\documentclass[twocolumn,floatfix,aps,prc,showpacs,superscriptaddress,preprintnumbers,amsmath,amssymb]{revtex4}

\usepackage{graphicx}
\usepackage{dcolumn}
\usepackage{bm}

\begin{document}


\title{$J/\psi$ Production in Au-Au Collisions at $\sqrt{s_{NN}}$ = 200 GeV}

\newcommand{\abilene}{Abilene Christian University, Abilene, TX 79699, USA}
\newcommand{\acadsin}{Institute of Physics, Academia Sinica, Taipei 11529, Taiwan}
\newcommand{\banaras}{Department of Physics, Banaras Hindu University, Varanasi 221005, India}
\newcommand{\barc}{Bhabha Atomic Research Centre, Bombay 400 085, India}
\newcommand{\bnl}{Brookhaven National Laboratory, Upton, NY 11973-5000, USA}
\newcommand{\caucr}{University of California - Riverside, Riverside, CA 92521, USA}
\newcommand{\ciae}{China Institute of Atomic Energy (CIAE), Beijing, People\'s Republic of China}
\newcommand{\cns}{Center for Nuclear Study, Graduate School of Science, University of Tokyo, 7-3-1 Hongo, Bunkyo, Tokyo 113-0033, Japan}
\newcommand{\columbia}{Columbia University, New York, NY 10027 and Nevis Laboratories, Irvington, NY 10533, USA}
\newcommand{\dapnia}{Dapnia, CEA Saclay, Bat. 703, F-91191, Gif-sur-Yvette, France}
\newcommand{\debrecen}{Debrecen University, H-4010 Debrecen, Egyetem t{\'e}r 1, Hungary}
\newcommand{\fsu}{Florida State University, Tallahassee, FL 32306, USA}
\newcommand{\gsu}{Georgia State University, Atlanta, GA 30303, USA}
\newcommand{\hiroshima}{Hiroshima University, Kagamiyama, Higashi-Hiroshima 739-8526, Japan}
\newcommand{\ihepprot}{Institute for High Energy Physics (IHEP), Protvino, Russia}
\newcommand{\isu}{Iowa State University, Ames, IA 50011, USA}
\newcommand{\jinrdubna}{Joint Institute for Nuclear Research, 141980 Dubna, Moscow Region, Russia}
\newcommand{\kaeri}{KAERI, Cyclotron Application Laboratory, Seoul, South Korea}
\newcommand{\kangnung}{Kangnung National University, Kangnung 210-702, South Korea}
\newcommand{\kek}{KEK, High Energy Accelerator Research Organization, Tsukuba-shi, Ibaraki-ken 305-0801, Japan}
\newcommand{\kfki}{KFKI Research Institute for Particle and Nuclear Physics (RMKI), H-1525 Budapest 114, POBox 49, Hungary}
\newcommand{\korea}{Korea University, Seoul, 136-701, Korea}
\newcommand{\kurchatov}{Russian Research Center ``Kurchatov Institute", Moscow, Russia}
\newcommand{\kyoto}{Kyoto University, Kyoto 606, Japan}
\newcommand{\labllr}{Laboratoire Leprince-Ringuet, Ecole Polytechnique, CNRS-IN2P3, Route de Saclay, F-91128, Palaiseau, France}
\newcommand{\lawllnl}{Lawrence Livermore National Laboratory, Livermore, CA 94550, USA}
\newcommand{\losalamos}{Los Alamos National Laboratory, Los Alamos, NM 87545, USA}
\newcommand{\lpc}{LPC, Universit{\'e} Blaise Pascal, CNRS-IN2P3, Clermont-Fd, 63177 Aubiere Cedex, France}
\newcommand{\lund}{Department of Physics, Lund University, Box 118, SE-221 00 Lund, Sweden}
\newcommand{\muenster}{Institut fuer Kernphysik, University of Muenster, D-48149 Muenster, Germany}
\newcommand{\myongji}{Myongji University, Yongin, Kyonggido 449-728, Korea}
\newcommand{\nagasaki}{Nagasaki Institute of Applied Science, Nagasaki-shi, Nagasaki 851-0193, Japan}
\newcommand{\newmex}{University of New Mexico, Albuquerque, NM, USA}
\newcommand{\nmsu}{New Mexico State University, Las Cruces, NM 88003, USA}
\newcommand{\ornl}{Oak Ridge National Laboratory, Oak Ridge, TN 37831, USA}
\newcommand{\orsay}{IPN-Orsay, Universite Paris Sud, CNRS-IN2P3, BP1, F-91406, Orsay, France}
\newcommand{\pnpi}{PNPI, Petersburg Nuclear Physics Institute, Gatchina, Russia}
\newcommand{\riken}{RIKEN (The Institute of Physical and Chemical Research), Wako, Saitama 351-0198, JAPAN}
\newcommand{\rkrbrc}{RIKEN BNL Research Center, Brookhaven National Laboratory, Upton, NY 11973-5000, USA}
\newcommand{\saispbstu}{St. Petersburg State Technical University, St. Petersburg, Russia}
\newcommand{\saopaulo}{Universidade de S{\~a}o Paulo, Instituto de F\'{\i}sica, Caixa Postal 66318, S{\~a}o Paulo CEP05315-970, Brazil}
\newcommand{\seoulnat}{System Electronics Laboratory, Seoul National University, Seoul, South Korea}
\newcommand{\stonybrkc}{Chemistry Department, Stony Brook University, SUNY, Stony Brook, NY 11794-3400, USA}
\newcommand{\stonycrkp}{Department of Physics and Astronomy, Stony Brook University, SUNY, Stony Brook, NY 11794, USA}
\newcommand{\subatech}{SUBATECH (Ecole des Mines de Nantes, CNRS-IN2P3, Universit{\'e} de Nantes) BP 20722 - 44307, Nantes, France}
\newcommand{\tenn}{University of Tennessee, Knoxville, TN 37996, USA}
\newcommand{\titech}{Department of Physics, Tokyo Institute of Technology, Tokyo, 152-8551, Japan}
\newcommand{\tsukuba}{Institute of Physics, University of Tsukuba, Tsukuba, Ibaraki 305, Japan}
\newcommand{\vandy}{Vanderbilt University, Nashville, TN 37235, USA}
\newcommand{\waseda}{Waseda University, Advanced Research Institute for Science and Engineering, 17 Kikui-cho, Shinjuku-ku, Tokyo 162-0044, Japan}
\newcommand{\weizmann}{Weizmann Institute, Rehovot 76100, Israel}
\newcommand{\yonsei}{Yonsei University, IPAP, Seoul 120-749, Korea}
\affiliation{\abilene}
\affiliation{\acadsin}
\affiliation{\banaras}
\affiliation{\barc}
\affiliation{\bnl}
\affiliation{\caucr}
\affiliation{\ciae}
\affiliation{\cns}
\affiliation{\columbia}
\affiliation{\dapnia}
\affiliation{\debrecen}
\affiliation{\fsu}
\affiliation{\gsu}
\affiliation{\hiroshima}
\affiliation{\ihepprot}
\affiliation{\isu}
\affiliation{\jinrdubna}
\affiliation{\kaeri}
\affiliation{\kangnung}
\affiliation{\kek}
\affiliation{\kfki}
\affiliation{\korea}
\affiliation{\kurchatov}
\affiliation{\kyoto}
\affiliation{\labllr}
\affiliation{\lawllnl}
\affiliation{\losalamos}
\affiliation{\lpc}
\affiliation{\lund}
\affiliation{\muenster}
\affiliation{\myongji}
\affiliation{\nagasaki}
\affiliation{\newmex}
\affiliation{\nmsu}
\affiliation{\ornl}
\affiliation{\orsay}
\affiliation{\pnpi}
\affiliation{\riken}
\affiliation{\rkrbrc}
\affiliation{\saispbstu}
\affiliation{\saopaulo}
\affiliation{\seoulnat}
\affiliation{\stonybrkc}
\affiliation{\stonycrkp}
\affiliation{\subatech}
\affiliation{\tenn}
\affiliation{\titech}
\affiliation{\tsukuba}
\affiliation{\vandy}
\affiliation{\waseda}
\affiliation{\weizmann}
\affiliation{\yonsei}
\author{S.S.~Adler}	\affiliation{\bnl}
\author{S.~Afanasiev}	\affiliation{\jinrdubna}
\author{C.~Aidala}	\affiliation{\bnl}
\author{N.N.~Ajitanand}	\affiliation{\stonybrkc}
\author{Y.~Akiba}	\affiliation{\kek} \affiliation{\riken} 
\author{J.~Alexander}	\affiliation{\stonybrkc}
\author{R.~Amirikas}	\affiliation{\fsu}
\author{L.~Aphecetche}	\affiliation{\subatech}
\author{S.H.~Aronson}	\affiliation{\bnl}
\author{R.~Averbeck}	\affiliation{\stonycrkp}
% ...
\collaboration{PHENIX Collaboration}  \noaffiliation

\date{\today}

\begin{abstract}

First results on charm quarkonia production in heavy ion collisions at
the Relativistic Heavy Ion Collider (RHIC) are presented. The yield of
$J/\psi$'s measured in the PHENIX experiment via electron-positron
decay pairs at mid-rapidity for Au-Au reactions at $\sqrt{s_{NN}}$ =
200 GeV are analyzed as a function of collision centrality. For this
analysis we have studied 49.3 million minimum bias Au-Au reactions. We
present the $J/\psi$ invariant yield $dN/dy$ for peripheral and
mid-central reactions. For the most central collisions where we
observe no signal above background, we quote 90\% confidence level
upper limits. We compare these results with our $J/\psi$ measurement
from proton-proton reactions at the same energy.  We find that our
measurements are not consistent with models that predict strong
enhancement relative to binary collision scaling.

\end{abstract}


\maketitle

\section{Introduction}

Lattice Quantum Chromodynamics (QCD) calculations indicate that there is a  
transition of nuclear matter from confined to deconfined quarks and gluons 
at a temperature of order $T_{c}=170$ MeV.  
Characteristic of this deconfined state of matter is the dynamic screening 
of the long-range confining potential of QCD.  Color screening is predicted to 
reduce the attraction between heavy quark-antiquark pairs, and thus leads to a 
decrease in the ratio of hidden charm and beauty (quarkonia) to open charm 
and beauty~\cite{matsui,latticeQCD}.
Thus, one expects a suppression of quarkonium states 
($J/\psi, \psi', \chi_{c}, \Upsilon(1s,2s,3s)$) depending on their binding energy and 
the temperature of the surrounding system.

In relativistic heavy ion collisions a state of a deconfined thermalized 
quark-gluon plasma may be created.  
Measurements in Pb-Pb reactions at 
$\sqrt{s_{NN}}$ = 17.3 GeV by the NA50 experiment~\cite{na50} show a 
suppression of heavy quarkonia production relative to ``normal'' nuclear 
absorption, the dissociation of $c\overline{c}$ pairs by interactions with 
the nucleons into separate quarks that eventually hadronize into 
$D$ mesons~\cite{leitch, arleo, dk, fujii}.  
This suppression has been interpreted in the context of color 
screening in a quark-gluon plasma~\cite{dk,blaizot}, 
additional absorption with co-moving 
hadrons~\cite{wong,capella}, and multiple scattering between the 
charm quarks and the surrounding medium~\cite{qiu,chaudhuri}.

At RHIC energies, where of order 10 $c\overline{c}$ pairs are produced in 
central Au-Au reactions~\cite{phx-elec,ralf}, some models predict an enhancement of 
heavy quarkonia due to $c\overline{c}$ coalescence in a quark-gluon 
plasma~\cite{thews}, detailed balance of 
$D + \overline{D} \leftrightarrow J/\psi + X$~\cite{rapp}, and/or 
statistical $J/\psi$ production~\cite{pbm}.  
In addition, at RHIC energies initial state effects of shadowing and possible 
parton saturation may play a role in initial charm production~\cite{hammon}.
Disentangling these competing effects will require a systematic study of yields 
of various quarkonium states in different colliding systems (proton-proton, 
proton (or deuteron)-ion, and ion-ion) and over a wide kinematic range in 
terms of transverse momentum and $x_{F}$.

We report here the first results on $J/\psi$ production via electron-positron
decay pairs at mid-rapidity from Au-Au
collisions at $\sqrt{s_{NN}}$ = 200 GeV from data taken during Run-2 at 
RHIC in 2001. 
For peripheral and mid-central Au-Au collisions, we present the most 
probable yield values, while for central reactions, 
we observe no signal above background and thus quote 90\% confidence level 
upper limits on $J/\psi$ production.

\section{PHENIX Experiment}
The PHENIX experiment is specifically designed to make use of high luminosity 
ion-ion, proton-ion, and proton-proton collisions at the Relativistic Heavy 
Ion Collider to sample rare physics probes including the $J/\psi$ and other 
heavy quarkonium states.  
The PHENIX experiment includes two central rapidity spectrometer arms, each covering
the pseudo-rapidity range $|\eta|<0.35$
and an interval of 90 degrees in azimuthal angle $\phi$.  
The spectrometers are comprised from the inner radius outward of a Multiplicity
and Vertex Detector (MVD), Drift Chambers (DC), Pixel Pad Chambers (PC),
Ring Imaging Cerenkov Counters (RICH), Time-of-Flight Scintillator Wall (TOF),
Time Expansion Chambers (TEC), and two types of Electromagnetic Calorimeters 
(EMC).
This combination of detectors allows for the clean identification of
electrons over a broad range in transverse momentum.  
Further details of the detector design and performance are given 
in~\cite{phenix-nim}.

The Au-Au event centrality is estimated using the combined data 
from our Beam-Beam Counters (BBC) and Zero Degree Calorimeters (ZDC).  
While the ZDCs measure forward neutrons that result from fragmentation of the 
colliding nuclei, the BBCs are sensitive to charged particles produced in
the collisions.
Together, both detectors yield information on the impact parameter of 
the nuclear reaction~\cite{zdc}.
These observables, combined with a Glauber model for the nuclear geometry, 
allow us to determine different collision geometry categories, referred
to as centrality ranges~\cite{ppg014}.

For the analysis presented here, the electron (positron) momentum and charge 
sign is determined from tracking using the DC and the PC 
and then projecting back through the PHENIX axial magnetic field to 
the collision point determined by the BBC~\cite{jeff}.
The momentum resolution achieved is $\delta p/p = 0.7\% \oplus 
1.0\%~\times p$~(in GeV/c).
Electrons are cleanly separated from the large background of charged pions 
and kaons by associating the tracks with at least three active photomultiplier 
tubes in the RICH~\cite{rich}.  
In addition, we compare the track momentum ($p$) to the energy ($E$) measured in the 
Electromagnetic Calorimeter.  
The $E/p$ ratio is used to further reduce the pion contamination in the 
electron sample.  
Pions typically deposit only a fraction of their energy in the calorimeter
whereas electrons deposit all of their energy.  
These selections are augmented by requiring that the calorimeter shower 
position and time-of-flight information agree with the track projection.  
Thus, we obtain a clean sample of electron and positron candidates with less than
5\% contamination.  

\section{Data Selection and Triggers}

The Au-Au data at $\sqrt{s_{NN}}=200$ GeV used in this analysis were recorded 
during Run 2 at RHIC in the fall of 2001.
For our ``minimum bias'' Au-Au event selection, we use a Level-1 trigger that 
requires a coincidence between our BBCs.  
We place an additional offline requirement of at least one forward neutron 
in each of our ZDCs to remove beam related backgrounds. 
Our ``minimum bias'' sample includes 92\% of the 6.9 barn Au-Au 
inelastic cross section~\cite{ppg014}.  
We further restrict our analysis to 90\% of the inelastic cross section
to remove a small remaining contribution from beam related background events.

We observed a Au-Au inelastic collision rate that increased 
during the running period from 100 to 1200 Hz.
The Level-2 triggers are implemented in a personal computer-based farm 
with 30 processors in Run 2, as part of the PHENIX Event 
Builder~\cite{phenix-nim}.
The Level-2 $J/\psi$ trigger algorithm identified electron candidates by
starting with rings in the RICH 
and then searching for possible matching showers in the EMC. 
The EMC search window based on the RICH ring is
obtained from a lookup table, generated using Monte Carlo simulations of
single electrons. Possible matches were assumed to be electron
candidates, and the electron momentum was taken to be the EMC shower energy.
The invariant mass was calculated for all electron candidate
pairs within an event, regardless of the candidate's charge sign. 
If the invariant mass was higher than 2.2 $\rm{GeV/c^{2}}$,
the pair was accepted as a $J/\psi$ candidate, and the entire event was archived.
The Level-2 trigger provided a rejection factor of order 30 relative to our 
``minimum bias'' Level-1 trigger sample.

An additional offline requirement was imposed that the collisions 
have a z-vertex satisfying 
$|z|<30~\rm{cm}$ in order to eliminate collisions taking place near the 
PHENIX magnet.
After this selection, we have analyzed 25.9 million ``minimum bias''
Au-Au reactions as triggered by our BBC Level-1  Trigger.
In addition, from the high luminosity period of running, we also processed
23.4 million ``minimum bias'' events with our $J/\psi$ Level-2 
trigger.

\section{$J/\psi$ Signal Counting}

\begin{figure*}[h]
\includegraphics[width=0.8\linewidth]{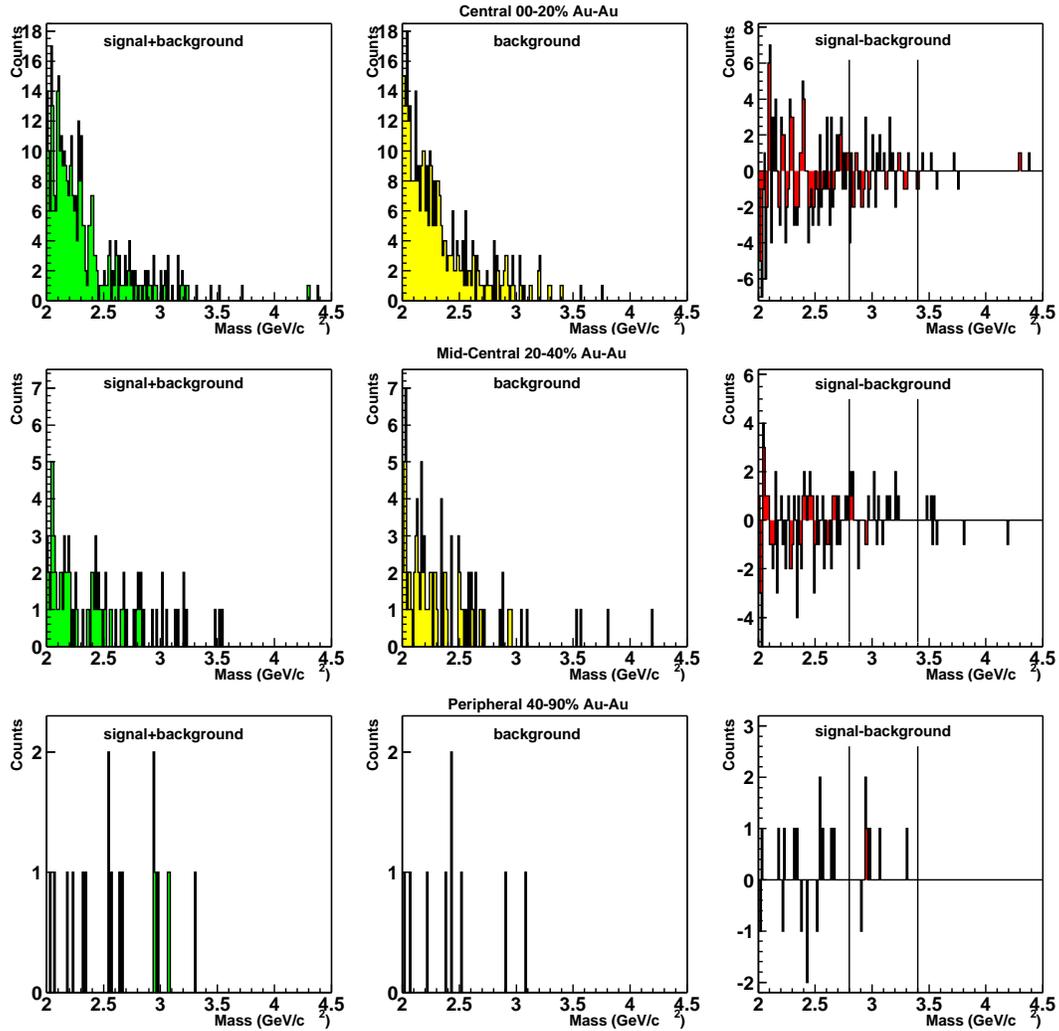}
\caption{\label{fig:mass}
(Color online) Dielectron invariant mass distribution in Au-Au reactions 
(top row: most central, 0-20\% central,
middle row: mid-central, 20-40\% central, and
bottom row: peripheral, 40-90\% central)
for unlike sign pairs containing signal+background (left column), like
sign pairs containing only background (center column) and the subtracted
difference (right column).}
\end{figure*}

For three exclusive centrality bins, 0-20\%, 20-40\%, and 40-90\% of the total
Au-Au cross section, we show the dielectron invariant mass distributions for 
unlike sign pairs ($e^{+}e^{-}$), like sign pairs ($e^{+}e^{+}$ or $e^{-}e^{-}$)
and the subtracted difference in Figure~\ref{fig:mass}.
The number of $J/\psi$ counts for each centrality range is determined from 
the number of signal counts above ``background'' within a fixed invariant 
mass window.  
The PHENIX acceptance and Level-2 trigger efficiencies are the same 
within a few percent for unlike sign pairs and like sign pairs 
in the $J/\psi$ mass region.  
Therefore, the sample of like sign pairs is a good representation with no
additional scale factor of the 
``background'' due to simple combinatorics.

In order to extract a $J/\psi$ signal strength, we employ a counting
method where we subtract from the number of unlike sign pairs the
number of like sign pairs in the mass window $2.8 < m < 3.4$
$\rm{GeV/c^{2}}$.  We have chosen a wide invariant mass window to be
consistent with the signal extraction method from our proton-proton
analysis~\cite{phenix-jpsi-pp}, and to limit our sensitivity to the
exact mass width value.  Although we expect from our Monte Carlo
studies a mass width of order 60 MeV, we cannot quantitatively verify
this even with our proton-proton data sample due to low statistics.
We note that in principle there is more information to be utilized in
the exact distribution of the candidates within the mass window.  
However, we have found that this does not add to the significance of
the result given the low counts and the lack of constraint on the
$J/\psi$ and background line shape.

\begin{table}
\caption{\label{tab:counts}
Statistical results for $J/\psi$ counts are shown for three
exclusive centrality ranges.  Shown are the number of unlike and like
sign counts within the mass window ($2.8 < m < 3.4~\rm{GeV/c^{2}}$).  
Also shown are the most likely signal value with the 68\% 
statistics confidence interval (for the peripheral and mid-central
cases), and the 90\% confidence level upper limits.}
\begin{ruledtabular}
 \begin{tabular}{ccccc}
Cen-    & Unlike Sign & Like Sign & Most Likely & 90\% \\ 
trality &   Counts    &   Counts  &   Signal    & C.L. \\ \hline
00-20\%  & 33  & 41  &  0  &  9.9  \\
20-40\% & 16  &  8  &  $8 ^{+4.8}_{-4.1}$  & 14.4  \\
40-90\% &  7  &  2  &  $5 ^{+3.1}_{-2.6}$  &  9.3 \\
 \end{tabular}
\end{ruledtabular}
\end{table}

\begin{figure}[tbh]
  \includegraphics[width=1.0\linewidth]{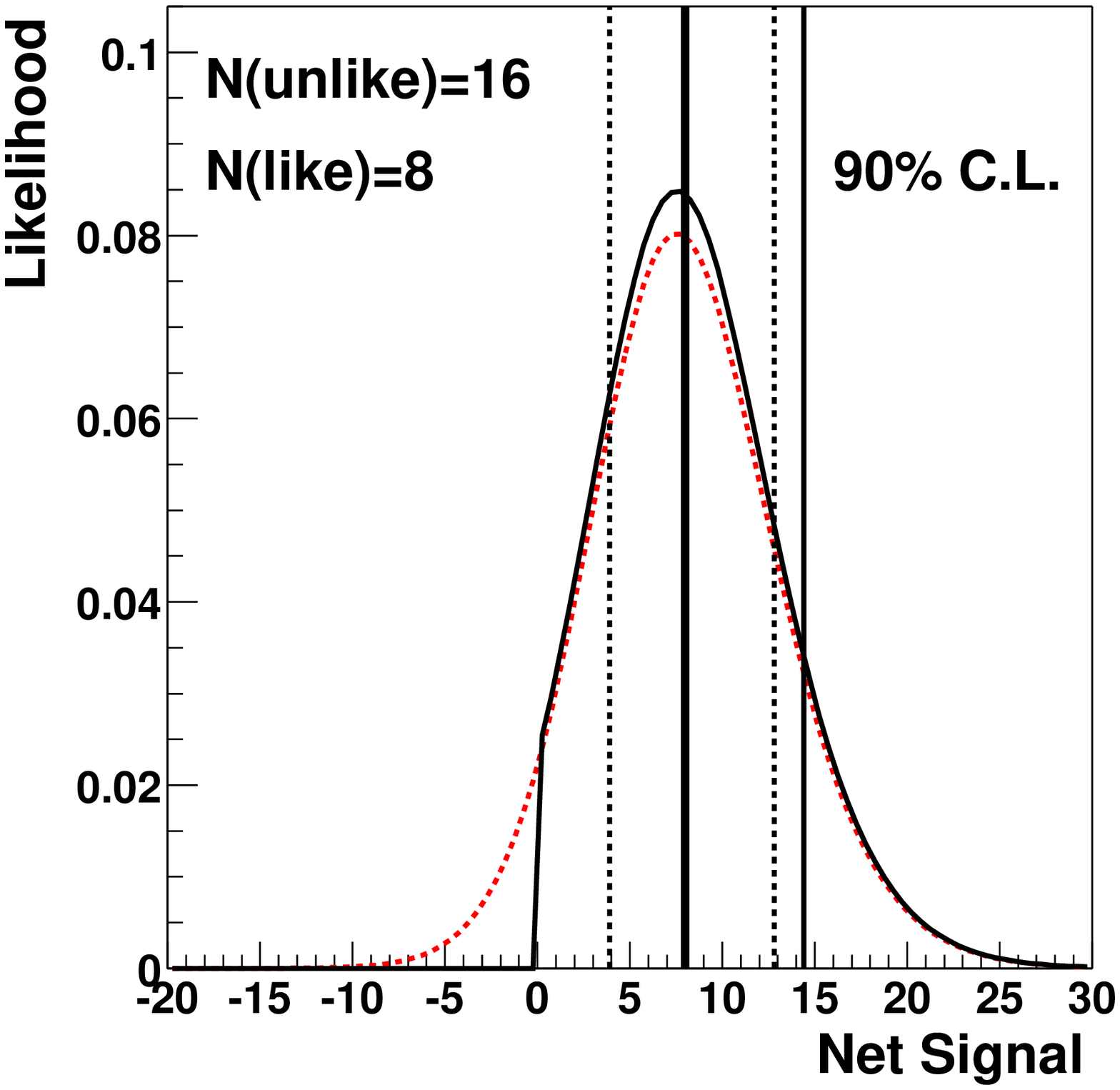}
  \caption{\label{fig:stats}
(Color online) The Poisson statistical likelihood distribution as a 
function of the expected net
signal.  The distributions are for the mid-central case of $N_{unlike}$=16 and $N_{like}$=8.  The
dashed curve is the likelihood distribution, and the black is after eliminating the unphysical
net signal less than zero and re-normalizing.  Vertical lines are shown to indicate the most
likely value (8), the 68\% confidence interval values, and the 90\% confidence level upper limit.}
\end{figure}

Table~\ref{tab:counts} shows the number of unlike and like sign counts 
within the mass window. 
For given observed counts of $N_l$ (like sign) and $N_u$ (unlike
sign), the likelihood $L(\nu_l,\nu_u)$ for the expectation values
$\nu_l$ and $\nu_u$ is given as
\begin{equation}
L(\nu_l,\nu_u) = {{\nu_{l}^{N_{l}}e^{-\nu_{l}}} \over {N_{l}!}} \times
{{\nu_{u}^{N_{u}}e^{-\nu_{u}}} \over {N_{u}!}}
\end{equation}
We then integrate $L(\nu_l,\nu_u)$ to give the likelihood $L(\nu_s)$
for the expectation value of the net signal counts $\nu_s = \nu_u - \nu_l$
\begin{equation}
L(\nu_s) = \int^{\infty}_{0} \int^{\infty}_{0} L(\nu_l,\nu_u) \delta(\nu_s - \nu_u + \nu_l)d\nu_l d\nu_u
\end{equation}
We show the likelihood distribution $L(\nu_s)$ for the mid-central (20\%-40\%)
events in Figure~\ref{fig:stats}. Although $L(\nu_s)$ is normalized such 
that $\int_{-\infty}^\infty L(\nu_s) d\nu_s = 1$, it has a non-zero 
probability for negative expected net signal value ($\nu_s < 0$).
Since the unlike sign contains $signal + background$, and the like sign 
contains only $background$, the only physically allowed values for $\nu_s$ are
greater than or equal to zero.
Thus, we remove the probability range corresponding to $\nu_s < 0$, and re-normalize the 
remaining probability integral to one~\cite{pdg}, as shown in the same Figure~\ref{fig:stats}.
We then determine for each centrality the 90\% confidence level upper 
limit, and the 68\% confidence interval around the most likely value for 
the peripheral and mid-central ranges.  
These values are shown in Table~\ref{tab:counts}.

Since the net signal is negative for the 0-20\% central event class, we can 
only quote a 90\% confidence level upper limit.  
Also, even for the 20-40\% and 40-90\% centrality classes, the signal observed 
is not significant at the two standard deviation level and thus we also show 
90\% confidence level upper limits for completeness.
The limited statistical significance of the results is clear from the mass 
distributions shown in Figure~\ref{fig:mass}.

In the intermediate mass region below the $J/\psi$,
$2.0 < m < 2.8~\rm{GeV/c^{2}}$, 
the shapes and absolute yield of like sign and unlike sign
dielectron mass distributions are well reproduced by an event mixing
method within a few percent. This indicates that most of the
dielectron pairs are from uncorrelated electron and positron
candidates. They are originating from Dalitz decays, photon conversions,
open charm/beauty semi-leptonic decays, and a small contamination of
mis-identified hadrons. The unlike sign pairs should also have a component
from semi-leptonic decays of charm and anti-charm pairs, but its
contribution is less than the statistical uncertainty of the data~\cite{nagle-qm}.
A complete analysis of the intermediate mass region will be presented elsewhere~\cite{hemmick}.

Our method of measurement does not separate contributions from decay 
feed-down from other states such as the $\chi_{c}$ and $\psi'$,
and thus our $J/\psi$ counts include these feed down sources.  
Our acceptance and efficiency is identical for the resulting $J/\psi$ from these
decays as for prompt $J/\psi$, and thus they enter our signal weighted simply by their relative production and
branching fraction into $J/\psi~+~X$.
Another contribution may result from $B$ meson decays into $J/\psi$.
However, if we assume a $b\overline{b}$ cross section in proton-proton 
reactions $\sigma_{b\overline{b}} \approx 2-5~\mu \rm{b}$~\cite{jaroschek} 
and that beauty production scales with binary collisions, 
we would expect a contribution of order 1-4\%
relative to primary $J/\psi$ or $J/\psi$ from $\chi_{c}$.  This percentage
contribution calculation assumes that the primary 
$J/\psi$ also scales with binary collisions.
If primary $J/\psi$ are substantially suppressed, the $B$ meson decay
contribution would constitute a larger fraction of our measured $J/\psi$,
especially at higher $p_{T}$.

It should also be noted that some signal in the unlike sign pairs from 
correlated open charm 
$c\overline{c} \rightarrow D (\rightarrow e^{+} + X) + \overline{D} (\rightarrow e^{-} + X)$ 
will contribute in our mass window.  
Assuming binary scaling in charm production with a proton-proton cross
section $\sigma_{c\overline{c}} \approx 650~\mu \rm{b}$~\cite{ralf},
this contribution in the $J/\psi$ mass region 
is estimated to be about 0.1 events in the 0-20\%, 0.05
events in the 20-40\%, and 0.02 events in the 40-90\% centrality bins.

\section{$J/\psi$ Yield Calculation}

We quote our results as the branching fraction of  
$J/\psi \rightarrow e^{+}e^{-}$ (B=$5.93 \pm 0.10 \times 10^{-2}$~\cite{pdg}) 
times the invariant yield at mid-rapidity $dN/dy|_{y=0}$.
We calculate this quantity for three exclusive centrality ranges as detailed 
below.

\begin{widetext}
\begin{equation}
B~{{dN} \over {dy}}|_{y=0} = {{N_{J/\psi}} \over {N_{mb-evt}+(\epsilon_{lvl2-eff}} \times N_{lvl2-evt})} \times {{1} \over {\Delta y}} \times {{1} \over {\epsilon_{acc-eff} \times \epsilon_{cent}}}
\end{equation}
\end{widetext}

The number of signal counts $N_{J/\psi}$ from both the ``minimum bias'' and 
Level-2 triggered event samples are shown in Table~\ref{tab:counts}.
The number of events from the ``minimum bias'' sample $N_{mb-evt}$ is 25.9 million Au-Au events.
The number of effective events sampled by the Level-2 trigger is 
$(\epsilon_{lvl2-eff} \times N_{lvl2-evt})$, which is the Level-2 trigger
efficiency times the number of events processed by the Level-2 trigger, 23.4 million Au-Au events.
This formulation appropriately weights the two data samples by the expected number of 
$J/\psi$.

The efficiency of the Level-2 trigger $\epsilon_{lvl2-eff}$ was determined by running the trigger
algorithm on simulated $J/\psi$ and
carrying out a full offline reconstruction of the resulting electron-positron decay pair.
The efficiency was calculated via counting the fraction of successfully
reconstructed $J/\psi$ events that were also found by the trigger. In these
trigger simulations, the channel-by-channel calibrations for the RICH and
EMC were used to convert the simulated signals into realistic values 
representative of a specific period in the run, before passing them to the 
Level-2 trigger. 

The overall $J/\psi$ trigger efficiency from the trigger simulations
was $\epsilon_{lvl2-eff}$ = 0.75 $\pm$ 0.04. The systematic error was
determined by studying the dependence of the trigger efficiency on the
collision vertex position, the assumed $J/\psi$ transverse momentum
and rapidity distribution, collision centrality, and the period
of the run from which the channel-by-channel calibrations were taken. 
After evaluating all of the above dependencies, we assign a
5\% systematic error to the $J/\psi$ trigger efficiency. 

The efficiency result from the trigger simulations was confirmed using
real data in two ways. First, the minimum bias data sample in Au-Au
collisions was analysed to calculate the $J/\psi$ trigger
efficiency. This was done by taking the ratio of the events that
have an electron pair with invariant mass between 2.8 and 3.4 $\rm{GeV/c^{2}}$ that
fired the $J/\psi$ trigger to all of the events having electron pairs in
that invariant mass range.
The trigger efficiency estimate from this check is 0.67 $\pm$ 0.10~(stat).
Second, the triggers were run on a sample of 26 events from the proton-proton
data set that passed all of the $J/\psi$ cuts in the offline analysis and had
invariant masses between 2.8 and 3.4 $\rm{GeV/c^{2}}$. 
The Level-2 $J/\psi$ trigger accepted 19 of these events, yielding an estimate 
of $0.73 \pm ^{0.07}_{0.10}$ (stat), in very good agreement with the trigger 
simulation result. 
These results verify that the trigger performance is similar for real data and 
simulations.  

\begin{figure}[tbh]
  \includegraphics[width=1.0\linewidth]{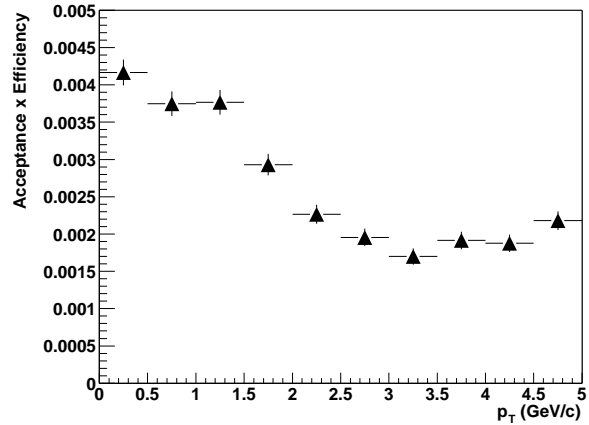}
  \caption{\label{fig:accept}
The PHENIX $J/\psi$ acceptance $\times$ efficiency as a 
function of the
$J/\psi$ transverse momentum is shown.  Most of the acceptance is with one lepton into each
of the two PHENIX central spectrometers.  This contribution peaks at $p_{T}=0$ and decreases
with increasing $p_{T}$.  The rise in the acceptance at high $p_{T}$ is from
contributions where both electron and positron are accepted into one of the PHENIX central
spectrometers.}
\end{figure}

The $J/\psi$ acceptance and efficiency $\epsilon_{acc-eff}$ is 
determined with a GEANT based Monte Carlo simulation of the PHENIX experiment.  
The detector response has been tuned to reproduce the resolution and 
performance of the real detector.   
The efficiency includes not only the tracking efficiency, but also the 
probability for passing all of the electron identification selection cuts.  
The electron identification efficiency determined by the Monte Carlo
is verified by a clean electron sample from conversion photons.
We also account for run by run efficiency changes by counting the relative 
number of reconstructed electrons and positrons per event in our data sample.
We show the PHENIX acceptance and efficiency as a function of transverse 
momentum in Figure~\ref{fig:accept}.

Since we do not have the statistics to determine the transverse momentum 
distribution of the $J/\psi$, we must employ a model for the $p_{T}$ 
dependence to determine an overall acceptance and efficiency.  
We use two different functional forms for the $p_{T}$ distributions to test
the model sensitivity of our acceptance.
We use an exponential in $p_{T}$ and an exponential in $p_{T}^{2}$ as
motivated by fits to $J/\psi$ data at lower energies~\cite{jpsi-lower}.  
The two models give similar acceptance values given a common $<p_{T}>$ 
value input.  
The largest uncertainty comes from the value of $<p_{T}>$ assumed. 
PHENIX has measured $J/\psi$ production in proton-proton reactions at 
$\sqrt{s}$ = 200 GeV and finds a 
$<p_{T}>$ = $1.80 \pm 0.23~(stat) \pm 0.16~(sys)$ GeV/c~\cite{phenix-jpsi-pp}.  
We use this value to determine our acceptance and efficiency averaged over all 
$p_{T}$.  The $J/\psi$ $<p_{T}>$ in Au-Au collisions may differ from that in proton-proton
reactions.
Therefore we vary the $<p_{T}>$ from 1.0 to 3.0 GeV/c to determine our model 
dependent systematic errors.
We assume that the $J/\psi$ rapidity distribution is flat over the range $-0.35 < y < 0.35$
where we measure.
The final value for the $J/\psi$ acceptance and efficiency is shown in 
Table~\ref{tab:eff}.
This acceptance and efficiency has a 20\% systematic error 
from uncertainties in matching the Monte Carlo to the detector response, a 
10\% systematic error from run-to-run variation corrections, and a 
$\pm ^{32}_{24}$\% systematic error from the uncertainty in the $<p_{T}>$.

\begin{table}
\caption{\label{tab:eff}
The $J/\psi$ acceptance $\times$ efficiency and the centrality 
dependent efficiency are shown for three exclusive Au-Au centrality event 
classes.}
\begin{ruledtabular} 
  \begin{tabular} {ccc} 
Centrality bin  &  $\epsilon_{acc-eff}$   &   $\epsilon_{cent}$  \\\hline
00-20\%           &  $0.0027 \pm ^{0.0009}_{0.0005} (sys)$  & $0.61 \pm 0.06~(sys)$  \\
20-40\%           &  $''$                                      & $0.78 \pm 0.08~(sys)$  \\
40-90\%           &  $''$                                      & $0.90 \pm 0.09~(sys)$  \\
  \end{tabular}
\end{ruledtabular}
\end{table}

Our tracking and electron identification efficiencies exhibit a centrality dependence
due to overlapping hits and energy contamination in the calorimeter.
We determine this dependence by embedding Monte Carlo $J/\psi$ into real data 
events of different centrality selections.  
The corresponding efficiency factor $\epsilon_{cent}$ varies from 56\% for the 
0-5\% most central events to 98\% for the 85-90\% most peripheral events.

The final values for the embedding efficiency in our wide centrality bins 
are sensitive to the true centrality dependence of the $J/\psi$ production.  
In order to estimate the systematic error due to this uncertainty we assume 
two different centrality dependence models: (1) binary collision scaling and 
(2) participant collision scaling.  
Within our centrality ranges, we find that these two models yield less than a 
5\% difference and we include this in our systematic error.
We assign an additional 10\% systematic error to account for uncertainties in 
the Monte Carlo embedding procedure.  The centrality dependent efficiency values 
are shown in Table~\ref{tab:eff}.  

In our $B~dN/dy$ calculation, we have added the systematic errors from 
all of the contributing factors in quadrature and find +35\% and -41\% total 
systematic error on the invariant yield in each of the centrality ranges.  
The dominant systematic error results from the uncertainty in the $<p_{T}>$ of 
the $J/\psi$ distribution.

\section{Results}

The $B~dN/dy|_{y=0}$ values for the three exclusive centrality selections are shown
in Table~\ref{tab:results1}.
We have calculated using a Glauber model~\cite{ppg014} the number of expected participating
nucleons $N_{part}$ and the number of expected binary collisions $N_{coll}$ for 
each centrality range. 
These results are shown in Table~\ref{tab:results2}, in addition to the 
$B~dN/dy|_{y=0}$ values divided by the expected number of binary collisions.

\begin{table}[tbh]
\caption{\label{tab:results1}
We show the statistically most likely $J/\psi$ invariant yield ($B~dN/dy|_{y=0}$) value and the
68\% confidence interval for peripheral (40-90\%) and mid-central (20-40\%) collisions.
We also show the 90\% confidence level upper limit and the systematic error on this limit 
for all three different centrality ranges of Au-Au collisions.}
\begin{ruledtabular} 
\begin{tabular} {ccc} 
                 &  \multicolumn{2}{c} {$B~dN/dy|_{y=0}~(\times 10^{-4})$} \\ \hline
Centrality       & Most Likely Value &   90\% C.L.U.L.      \\ \hline

00-20\%          & N.A. & $6.08 {+1.56}~(sys)$\\

20-40\%          & $4.00 ^{+2.34}_{-2.01}~(stat) ^{+1.36}_{-1.60}~(sys)$ & $7.19 + {2.43} ~(sys)$\\

40-90\%          & $0.86 ^{+0.52}_{-0.44}~(stat) ^{+0.29}_{-0.35}~(sys)$ & $1.60 + {0.54} ~(sys)$\\
\end{tabular}
\end{ruledtabular}
\end{table}
\begin{table*}[bth]
\caption{\label{tab:results2}
We show the number of participating nucleons and the number of binary collisions for three
different centrality ranges of Au-Au collisions, and the associated systematic errors.
We show the statistically most likely value for the $J/\psi$ invariant yield ($B~dN/dy|_{y=0}$) divided
by the expected number of binary collisions for peripheral (40-90\%) and mid-central (20-40\%) collisions.
We also show the 90\% confidence level upper limit and the systematic error on this limit 
for all three different centrality ranges of Au-Au collisions.
The systematic error in the invariant yield per binary collision does not include 
the systematic error in the expected number of binary collisions.
This error contribution is negligible for the central and mid-central categories and would
increase the systematic error for the peripheral category by 6\%.}
\begin{ruledtabular} 
\begin{tabular} {ccccc} 
                 &  & & \multicolumn{2}{c} {$B~dN/dy|_{y=0}$ per binary collision ($\times 10^{-6}$)} \\
Centrality       & $N_{part}$  & $N_{coll}$ & Most Likely Value &   90\% C.L.U.L.      
\\ \hline
00-20\%          & $280 \pm 4$ & $779 \pm 75$ & N.A. & $0.78 {+0.20} ~(sys)$\\
20-40\%          & $140 \pm 5$ & $296 \pm 31$ & $1.35 ^{+0.79}_{-0.68}~(stat) ^{+0.46}_{-0.54}~(sys)$ & $2.43 +{0.82} ~(sys)$\\
40-90\%          & $34 \pm 3$  &  $45 \pm 7$  & $1.91 ^{+1.15}_{-0.97}~(stat) ^{+0.65}_{-0.77}~(sys)$ & $3.55 {+1.21} ~(sys)$\\
\end{tabular}
\end{ruledtabular}
\end{table*}

The PHENIX result for the $J/\psi$ invariant yield in proton-proton induced reactions 
at $\sqrt{s}$=200 GeV at mid-rapidity~\cite{phenix-jpsi-pp} is
\begin{widetext}
\begin{equation}
B~dN/dy|_{y=0} (pp) = 1.46 \pm 0.23~(stat) \pm 0.22~(sys) \pm 0.15~(abs) \times 10^{-6}
\end{equation}
\end{widetext}
The systematic error (abs) represents the uncertainty of the normalization of the total 
proton-proton invariant yield.

\begin{figure}[tbh]
  \includegraphics[width=1.0\linewidth]{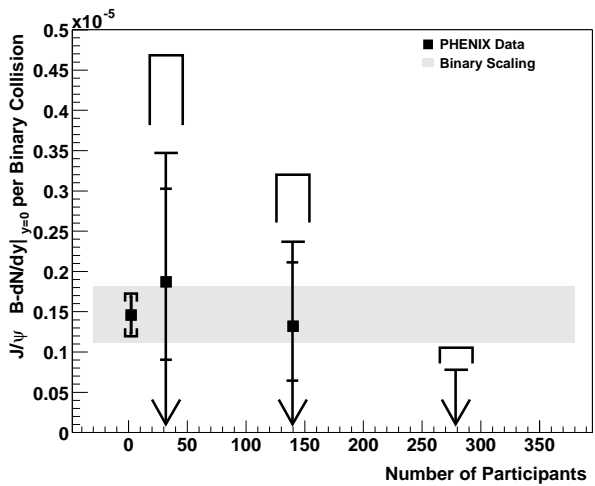}
  \caption{\label{fig:data}
The $J/\psi$ invariant yield per binary collision is shown 
for proton-proton reactions and three
exclusive centrality ranges of Au-Au reactions all at $\sqrt{s_{NN}}$ = 200 GeV.  
For the proton-proton reactions, we show the most likely value as a data point 
(square), the statistical error, and the 
estimated systematic errors as brackets.  
For the three Au-Au data points, we show as arrows the 90\% confidence level 
upper limits.  
The bracket above the limit includes the estimated systematic error on these limits.  
In the case of the peripheral and mid-central ranges, we also show, as a 
square marker, the statistically most likely value and as two horizontal 
dashes the 68\% confidence interval.  The gray band indicates binary scaling and the
width is the quadrature sum of the statistical and systematic error on our proton-proton data point.
For the Au-Au points, the systematic error in the invariant yield per binary collision does not include 
the systematic error in the expected number of binary collisions.
This error contribution is negligible for the central and mid-central categories and would
increase the systematic error for the peripheral category by 6\%.}
\end{figure}
We show in Figure~\ref{fig:data} the results from the three Au-Au 
centrality bins and the proton-proton data normalized per binary collision as 
a function of the number of participating nucleons.  
Note that for proton-proton reactions, there are two participating nucleons 
and one binary collision.

\section{Discussion}

Despite the limited statistical significance and systematic uncertainty of these
first $J/\psi$ results, we can address some important 
physics questions raised by the numerous theoretical 
frameworks in which $J/\psi$ rates are calculated.

\begin{figure}[tbh]
  \includegraphics[width=1.0\linewidth]{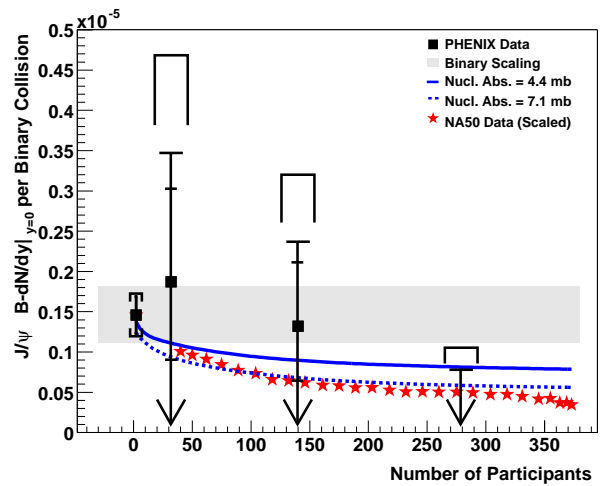}
  \caption{\label{fig:datamodelA}
(Color online) The $J/\psi$ invariant yield per binary collision is shown 
from proton-proton reactions and three
exclusive centrality ranges of Au-Au reactions all at $\sqrt{s_{NN}}$ = 200 GeV.
The lines are the theoretical expectations from ``normal'' nuclear absorption with 
$\sigma_{c\overline{c}-N}$ = 4.4 mb (solid curve) and 7.1 mb (dashed curve) cross section.  
The stars are the $J/\psi$ per binary
collision measured by the NA50 experiment at lower collision energy.  In order to compare the
shapes of the distribution, we have normalized the NA50 data to match the central value for
our proton-proton results.}
\end{figure}

We show in Figure~\ref{fig:datamodelA} binary scaling expectations as a 
gray band.  We also show a calculation of the suppression expected from ``normal'' 
nuclear absorption using a $\sigma_{c\overline{c}-N} = 4.4$ mb~\cite{na50pa} 
and $7.1$ mb~\cite{nagle,dk}.
A recent measurement in proton-nucleus collisions at lower energies~\cite{na50pa} 
favors the smaller absorption cross section, thus
underscoring the importance of measuring $J/\psi$ in proton(deuteron)-nucleus collisions at 
RHIC energies.
We also show the NA50 suppression pattern relative to binary 
scaling~\cite{na50}, normalized to match our proton-proton data point at 
200 GeV.  
The data disfavor binary scaling, while they are consistent with ``normal'' 
nuclear absorption alone and also the NA50 suppression pattern measured at 
lower energies, within our large statistical errors.

\begin{figure}[tbh]
  \includegraphics[width=1.0\linewidth]{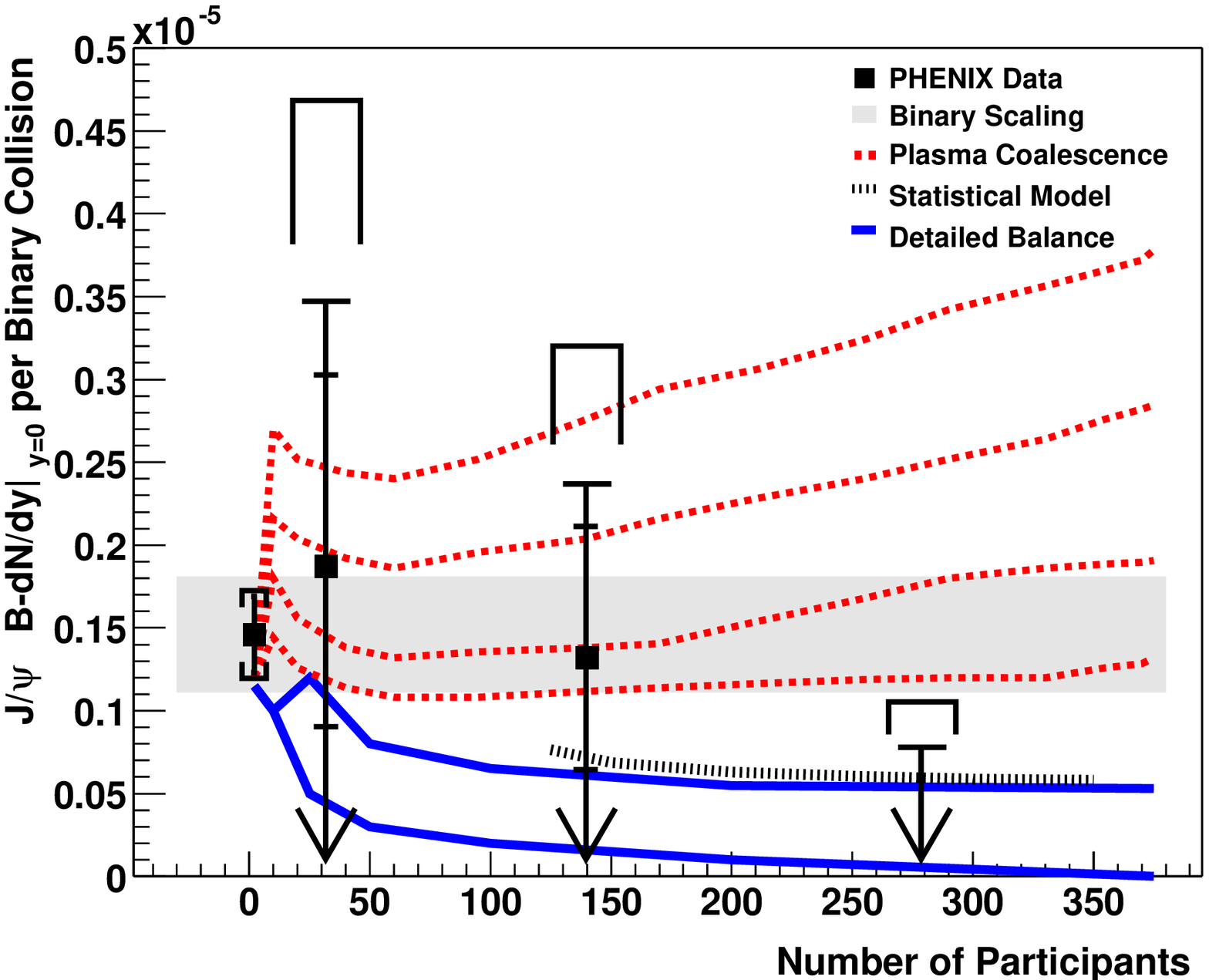}
  \caption{\label{fig:datamodelB}
(Color online) The $J/\psi$ invariant yield per binary collision is shown 
from proton-proton reactions and three
exclusive centrality ranges of Au-Au reactions all at $\sqrt{s_{NN}}$ = 200 GeV.
The lowest curve is a calculation including ``normal'' nuclear absorption in addition
to substantial absorption in a high temperature quark-gluon plasma~\cite{rapp}.  The curve above
this is including backward reactions that recreate $J/\psi$.  The statistical model~\cite{pbm}
result is shown as a dotted curve for mid-central to central collisions just above that.
The four highest dashed curves are from the plasma coalescence model~\cite{thews} for a temperature
parameter of $T$ = 400 MeV and charm rapidity widths of $\Delta y$ = 1.0, 2.0, 3.0, 4.0, from the
highest to the lowest curve respectively.
}
\end{figure}

One model calculation~\cite{rapp} including just the ``normal'' nuclear and plasma 
absorption components at RHIC energies is shown in Figure~\ref{fig:datamodelB}.
The higher temperature ($T$) and longer time duration of the system at RHIC lead 
to a predicted larger suppression of $J/\psi$ relative to binary collision 
scaling.  
This specific model~\cite{rapp}, and in general this class of 
models~\cite{satz,blaizot}, cannot be ruled out at this time due to our null
result (90\% confidence level upper limit) for the most central collisions.

Many recent theoretical calculations also include the possibility for 
additional late stage re-creation or coalescence of $J/\psi$ states.  
In ~\cite{rapp}, they include both break-up and creation reactions 
$D + \overline{D} \leftrightarrow J/\psi + X$.  
At the lower fixed target CERN energies, this represents a very small 
contribution due to the small charm production cross section.  
However, at RHIC energies, where in central Au-Au collisions there are of 
order 10 $c\overline{c}$ pairs produced, the contribution is significant.
The sum of the initial production, absorption, and re-creation as
shown in Figure~\ref{fig:datamodelB} is also consistent with our experimental data.  

A different calculation~\cite{thews} assumes the formation of a quark-gluon 
plasma in which the mobility of heavy quarks in the deconfined region leads to
increased $c\overline{c}$ coalescence.  
This leads to a very large enhancement of $J/\psi$ production at RHIC energies 
for the most central reactions.  
The model considers the plasma temperature ($T$) and the rapidity width 
($\Delta y$) of charm quark production as input parameters.  
Shown in Figure~\ref{fig:datamodelB} are the calculation results for 
$T$ = 400 MeV and $\Delta y$ = 1.0, 2.0, 3.0, 4.0.  
The narrower the rapidity window in which all charm quarks reside, the larger 
the probability for $J/\psi$ formation.  
$\Delta y=1.0$ is consistent with the three dimensional spherically symmetric 
thermal distribution, and results in a charm yield at midrapidity that is 
inconsistent with the PHENIX preliminary charm yield as determined from 
single electron measurements~\cite{ralf}.  
$\Delta y=4.0$ is consistent with expectations from factorized QCD and 
PYTHIA with CTEQ5L structure functions~\cite{phx-elec}.
All of these parameters within this model predict a $J/\psi$ enhancement 
relative to binary collisions scaling, which is disfavored by our data.

Another framework for determining quarkonia yields is to assume a statistical 
distribution of charm quarks that may then form quarkonia.  
A calculation assuming thermal, but not chemical equilibration~\cite{pbm} 
is shown in Figure~\ref{fig:datamodelB}, and is also consistent with our data.  

Significantly larger data sets are required to address the various models 
that are still consistent with our first measurement.  
Key tests will be the $p_{T}$ and $x_{F}$ dependence of the $J/\psi$ yields, 
and how these compare with other quarkonium states such as the $\psi'$.

\section{Summary}

PHENIX has shown first results on $J/\psi$ production in Au-Au collisions at 
$\sqrt{s_{NN}}$ = 200 GeV at mid-rapidity as measured via electron-positron 
pairs.  We find that models that predict $J/\psi$ enhancement relative to 
binary collision scaling are disfavored, while we cannot discriminate between
various scenarios leading to suppression relative to binary scaling.

This first measurement from PHENIX will be followed with high statistics 
measurements in both the electron channel at midrapidity and at forward and 
backward rapidities in the PHENIX muon spectrometers.  
Such measurements are expected in the next few years and will address the full range of heavy 
quarkonia production and evolution models.

\begin{acknowledgments}


We thank the staff of the Collider-Accelerator and Physics
Departments at Brookhaven National Laboratory and the staff of
the other PHENIX participating institutions for their vital
contributions.  We acknowledge support from the Department of
Energy, Office of Science, Nuclear Physics Division, the
National Science Foundation, Abilene Christian University
Research Council, Research Foundation of SUNY, and Dean of the
College of Arts and Sciences, Vanderbilt University (U.S.A),
Ministry of Education, Culture, Sports, Science, and
Technology and the Japan Society for the Promotion of Science
(Japan), Conselho Nacional de Desenvolvimento Cient\'{\i}fico
e Tecnol{\'o}gico and Funda\c c{\~a}o de Amparo {\`a} Pesquisa
do Estado de S{\~a}o Paulo (Brazil), Natural Science
Foundation of China (People's Republic of China), IN2P3/CNRS
and Commissariat a l'Energie Atomique (France),
Bundesministerium fuer Bildung und Forschung, Deutscher
Akademischer Austausch Dienst, and Alexander von Humboldt
Stiftung (Germany), Hungarian National Science Fund, OTKA
(Hungary), Department of Atomic Energy and Department of
Science and Technology (India), Israel Science Foundation
(Israel), Korea Research Foundation and Center for High Energy
Physics (Korea), Russian Academy of Science, Ministry of
Atomic Energy of Russian Federation, Ministry of Industry,
Science, and Technologies of Russian Federation (Russia), VR
and the Wallenberg Foundation (Sweden), the U.S. Civilian
Research and Development Foundation for the Independent States
of the Former Soviet Union, the US-Hungarian NSF-OTKA-MTA, the
US-Israel Binational Science Foundation, and the 5th European
Union TMR Marie-Curie Programme.

\end{acknowledgments}



\end{document}